\documentstyle[emulateapj,epsf,apjfonts]{article}

\slugcomment{Submitted to the Astrophysical Journal}

\begin{document}

\title{Fundamental Properties and Distances of LMC Eclipsing Binaries: III. 
EROS~1044{\footnote{Based on observations with the NASA/ESA Hubble Space 
Telescope, obtained at the Space Telescope Science Institute, which is operated 
by the Association of Universities for Research in Astronomy, Inc. under NASA 
contract No. NAS5-26555.}}}

\author{Ignasi Ribas\altaffilmark{2}, Edward L. Fitzpatrick, Frank P.  
Maloney, Edward F. Guinan} 
\affil{Department of Astronomy \& Astrophysics, Villanova University,
Villanova, PA 19085}

\author{Andrzej Udalski}
\affil{Warsaw University Observatory}

\altaffiltext{2}{Currently at Departament d'Astronomia i Meteorologia,
Universitat de Barcelona, Av. Diagonal, 647, E-08028 Barcelona, Spain}

\begin{abstract}

We present results from a detailed analysis of a third eclipsing binary
(EB) system in the Large Magellanic Cloud, EROS~1044 ($\sim$B2 IV--V +
$\sim$B2 III--IV).  Our study combines the ``classical'' EB study of light
and radial velocity curves with detailed modeling of the observed spectral
energy distribution, and yields an essentially complete picture of the
stellar properties of the system and a determination of its distance.  
The observational data exploited include optical photometry, space-based
UV spectroscopy, and UV/optical spectrophotometry.  The advantages of our
technique include numerous consistency checks and, in the case of the
distance determinations, the absence of zero point uncertainties and
adjustable parameters.  We find the EROS~1044 system to consist of a pair
of normal, mildly-evolved $\sim21000$ K stars, whose derived properties
are consistent with stellar evolution calculations.  The distance to the
system is $47.5\pm1.8$ kpc.  We discuss the implications of our results
for three EB systems (HV~2274, HV~982, and EROS~1044) on the general
distance to the Large Magellanic Cloud.

\end{abstract}


\section{Introduction}

This is the third in a series of papers presenting results from detailed
analyses of detached main sequence B-type eclipsing binary (EB) systems in
the Large Magellanic Cloud (LMC).  The main scientific goals of this
program are:  1) to determine an essentially complete description of the
stellar properties of each system (including temperature, mass,
luminosity, and radius); and 2) to measure a precise distance for each
system, from which the general distance to the LMC may be derived.  The
former goal allows for rigorous testing of the predictions of stellar
structure and evolution theory, while the latter addresses the important
role the LMC plays in the calibration of the cosmic distance scale and
seeks to reduce the uncertainty in the LMC's distance --- currently
10--15\% --- to a level of a few percent.

Our analysis is data-intensive, requiring precise photometry (yielding
light curves), high-resolution spectroscopy (yielding radial velocity
curves), and UV-optical spectrophotometry spanning the range
$\sim$1200--7000~\AA.  In addition to the derivation of fundamental
stellar quantities, the benefits of the analysis include numerous
consistency checks and, in the case of the distance determinations, the
absence of zero point uncertainties and adjustable parameters.  In fact,
the only parameter not explicitly determined directly is the extinction
ratio $R$ ($\equiv A(V)/E(B-V)$). (This quantity is, however, potentially
measurable --- although not with currently available data.)  Our results
do depend somewhat on stellar atmosphere models, although the main
implication of this is to restrict the temperature and evolutionary state
of the selected targets to a range within which the models are known to be
physically appropriate and to yield precise and accurate results.  This
dependence on model atmospheres leads to our focus on only detached,
mildly evolved, early-B EBs.

In our previous studies, we examined the LMC EB systems HV~2274 (Guinan et
al. 1998, hereafter ``Paper I''; Ribas et al. 2000a) and HV~982
(Fitzpatrick et al. 2002, ``Paper II'').  The apparent locations of these
systems in the LMC can be seen in Figure \ref{figLMC}.  For HV~2274,
located in the LMC's Bar, we find a distance of $47.0\pm2.2$ kpc
(corresponding to a distance modulus of $V_0-M_V = 18.36$ mag).  HV~982,
located somewhat north of the Bar and in the apparent vicinity of the 30
Doradus complex, yields a larger distance of 50.2$\pm$1.2 kpc
(corresponding to $V_0-M_V = 18.50$ mag). Given the uncertainties, these
results are apparently compatible but, when corrected for the most
commonly assumed orientation of the LMC, they predict distances to the
LMC's optical center which differ by nearly 5 kpc.  This might indicate a
large line-of-sight spread in the distribution of early-type LMC stars, or
perhaps a problem with the assumed LMC orientation.  This issue can only
be resolved by observation and analysis of other EB systems, located along
a variety of sightlines.

In this paper we apply our analysis to a third LMC EB system, EROS~1044,
and derive its stellar properties and distance. This system consists of
two mildly evolved $\sim$B2 stars, with an orbital period of 2.73 days,
and $V \simeq 15.2$. Preliminary determinations of distance and absolute
dimensions were discussed by Maloney et al. (2000, 2001). EROS~1044 was
discovered by the EROS microlensing project (Grison et al. 1995) and has
also been observed by the OGLE microlensing project (Udalski, Kubiak, \&
Szymanski 1997), with an OGLE designation of LMC\_SC7 254744. While EROS
and OGLE, as well as the MACHO program (Alcock et al. 1997), are directed
toward the discovery of compact objects in the halo of the Milky Way,
these programs have produced extremely valuable photometric archives of
variable stars in the LMC --- notably eclipsing binaries and Cepheid
variables.  These resources provide extensive masterlists from which
objects can be selected for further study. In the present case, EROS~1044
was originally selected from the EROS catalog based on the properties of 
its brightness and its light curve, indicating a detached system with 
deep and total eclipses.

In \S 2, we describe the data included in this study. In \S 3, the three
components of the analysis (involving the light curve, radial velocity,
curve and spectral energy distribution) are discussed and their individual
results presented. Some aspects of our results relating to the
interstellar medium towards EROS 1044 are described in \S 4, including an
indication of the relative location of the system within the LMC.  In \S
5, we show how the results of the analyses combine to yield a thorough
description of the stellar properties of EROS~1044, and we examine the
consistency of the results with stellar evolution theory.  Finally, in \S
6, we show how the distance to the system is derived from our analysis and
compare it with our previous results.

\section{The Data}

Three distinct datasets are required to carry out our analyses of the LMC
EB systems: precise differential photometry (yielding light curves),
high-resolution spectroscopy (yielding radial velocity curves), and
multiwavelength spectrophotometry (yielding temperature and reddening
information).  Each of these three are described briefly below.

As in our previous papers, the primary ($P$) and secondary ($S$)
components of the EROS~1044 system are defined photometrically and refer
to the hotter and cooler components, respectively.

\subsection{Optical Photometry} \label{secOPHOT}

CCD differential photometric observations of EROS~1044 were obtained by
both the EROS and OGLE projects. EROS data were secured with a 40-cm
telescope between 1991 and 1992, and consist of 511 measurements in the
blue ($\lambda$ $\sim$4900~\AA) and 542 measurements in the red ($\lambda$
$\sim$6700~\AA). Photometric indices were obtained via the technique of
point spread function fitting. The OGLE data were acquired between 1997
and 2000 with the 1.3-m Warsaw telescope at Las Campanas Observatory,
Chile, and consist of 410 $I$-band measurements. In this case, the
technique of Difference Image Analysis was employed to derive photometric
measurements, as described in Zebrun, Soszynski, \& Wozniak (2001).

\begin{figure*}
\plotone{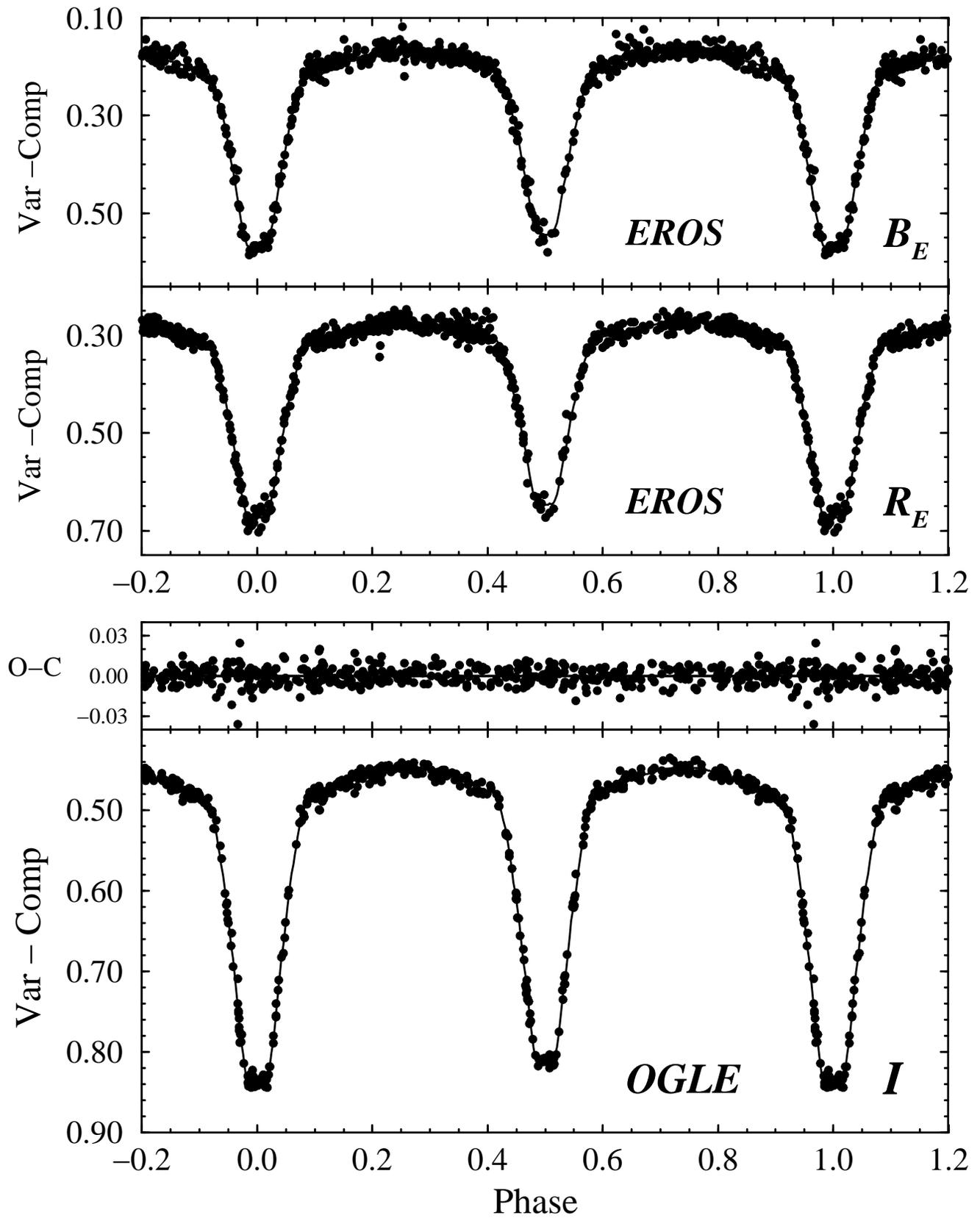}
\caption[f2.eps]{EROS $B_{\rm E}$, EROS $R_{\rm E}$, and OGLE $I$
light curves for EROS~1044 (filled circles) overplotted with the best fitting
model (solid curves). The residuals to the fit of the OGLE light curve
(``O--C'') are shown above. The parameters derived from the fit are listed 
in Table \ref{tabPARMS}. 
\label{figLC}}
\end{figure*}

The only available ephemeris for EROS 1044 comes from the EROS catalogue,
which indicates $P=2.727$ days. With a much longer time baseline now
available, we attempted to refine the orbital period and determine a new
reference epoch. We soon discovered what appears to be a discrepancy with
the Julian Dates provided in the EROS light curve database. Several tests
indicated a difference of about $0.125$ days ($=3$ hours) between the
expected times and the listed EROS times. This 3-hour difference is
suspicious and may result from the listing of Chilean local time (where
the observations were acquired) rather than UTC (Pritchard, private
communication). We therefore decided to ignore the EROS absolute time
scale altogether and base the period determination solely on the time
intervals within the OGLE and EROS observations. Thus, a new orbital
ephemeris was determined through simultaneous minimization of the
residuals of both EROS and OGLE light curves. This approach proved to be
successful (due, essentially to the long time span of the OGLE
observations) and yielded the following accurate ephemeris:  
\[\begin{array}{rcrcrc} 
T(\mbox{Min I})&=&{\rm HJD}2450868.2518&+&2.727125 & E \\
               & & 7& & 4 & \\
\end{array}\] 
\noindent The resultant light curves are well-defined, as shown in
Figure \ref{figLC}, with the EROS observations having somewhat greater
scatter than the OGLE data. The light curves exhibit primary and
secondary minima with nearly equal depths of $\sim$0.35 mag. The
portions of the light curve between eclipses show small brightness
variations presumably arising from the ellipticity of the stars and
reflection effects.

The OGLE database also contains V-band measurements for EROS~1044. While
the number of these ($\sim$35) is insufficient for our light curve
analysis, they nevertheless provide a measure of the out-of-eclipse V
magnitude for the EROS~1044 system, which is V$_{max} = 15.265\pm0.020$.

\subsection{UV Spectroscopy}

Radial velocity curves for EROS~1044 were derived from UV spectra obtained
by us between December 2000 and January 2001 with the Space Telescope
Imaging Spectrograph (STIS) of the {\it Hubble Space Telescope} (HST). The
observations were made using the FUV-MAMA detector, with the
52\arcsec$\times$0.5\arcsec\ aperture and G140M grating. We secured nine
spectra of EROS~1044 -- near orbital quadratures -- covering the
wavelength range 1294--1348~\AA, with a spectral resolution of
$\lambda/\Delta\lambda \simeq 15500$, and a S/N of $\sim$20:1.  The plate
scale of the data is 0.05~\AA~pix$^{-1}$ (12~km~s$^{-1}$~pix$^{-1}$) and
there are 1.7 pixels per resolution element. Identical instrumental setups
were used for all observations. The exposure time per spectrum was
$\sim$2200 s, sufficiently short to avoid significant radial velocity
shifts during the integrations. All the EROS~1044 observations were
reduced using the standard HST pipeline reduction software. The dataset
identifiers are O665A1010, O665A1020, O665A2010, O665A2020, O665A3010,
O665A4010, O665A5010, O665A6010, and O665A7010.

\begin{figure*}
\plotone{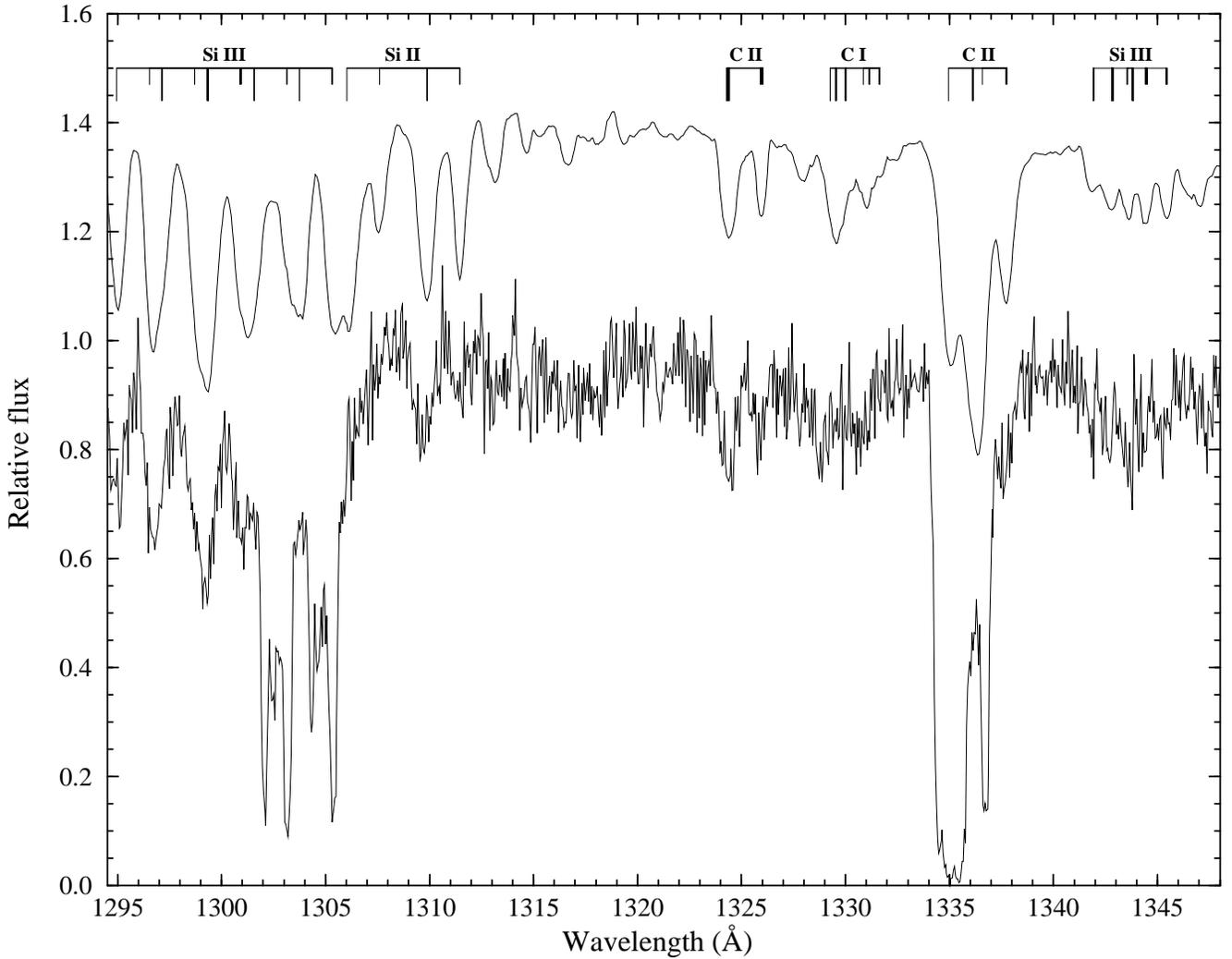}
\caption[f3.eps]{Observed spectrum of EROS 1044 obtained with
HST/STIS at binary phase 0.7911. The velocities of the primary and
secondary components at this phase are 459.9~km~s$^{-1}$ and
96.9~km~s$^{-1}$, respectively.  A double-line synthetic model is shown
for comparison, with prominent stellar lines labeled (see \S 2.2).  The
features of the more luminous secondary star are indicated by the long
tick marks, and those of the primary star by the short tick marks.  The
strong absorption complexes near 1302.5 \AA, 1305 \AA, and 1336
\AA\ are interstellar in origin, arising in both Milky Way and LMC gas,
and result from O I ($\lambda$ 1302.2), Si II ($\lambda$ 1304.4), and C
II ($\lambda\lambda$ 1334.5, 1335.7), respectively.
\label{figSPEC}}
\end{figure*}

For illustration, one of the HST/STIS spectra of EROS~1044 (O665A2020),
taken at phase 0.79, is shown in Figure \ref{figSPEC}. For reference, a
synthetic model is shown above the STIS spectrum, with prominent stellar
features identified. The model was constructed by Doppler shifting and
combining two synthetic spectra, each computed using R.L Kurucz's {\it
ATLAS9} atmosphere models and I. Hubeny's spectral synthesis program {\it
SYNSPEC}. The determination of the appropriate stellar
parameters\footnote{The synthetic spectrum for the distorted secondary
star was computed by assuming a constant mean temperature over the surface
even though differences of $\sim$3\% between the polar and the equatorial
temperatures are expected.} discussed in \S 3, and all the stellar
properties are summarized in \S 5.  The strong absorption complexes near
1302.5 \AA, 1305 \AA, and 1336 \AA\ are interstellar in origin, arising in
both Milky Way and LMC gas, and result from O I ($\lambda$ 1302.2), Si II
($\lambda$ 1304.4), and C II ($\lambda\lambda$ 1334.5, 1335.7),
respectively.

Radial velocities were measured from the HST/STIS data using the
``spectral disentangling'' technique. As discussed in Paper II, this
technique assumes that an observed double-lined spectrum is a linear
combination of two single-lined spectra (one per component) at different
relative velocities (determined by the orbital phase at the time of
observation). The analysis yields the two single-line spectra and the set
of relative velocities for each observed spectrum by considering the
entire dataset simultaneously.  The practical implementation of spectral
disentangling in our case was done through the Fourier disentangling code
{\it KOREL} developed by Hadrava (1995, 1997). The spectral data were
prepared for analysis with {\it KOREL} by rebinning to 1024 equal radial
velocity bins.  Data in the regions affected by interstellar absorption
were masked out by setting the normalized flux between 1301.6--1305.8~\AA\
and 1334.0--1338.3~\AA\ to unity. A large number of {\it KOREL} runs from
different starting points were carried out to explore the parameter space
and make estimations of the uncertainties.

The radial velocities produced by the {\it KOREL} analysis are not
heliocentric, but rather are referred to the barycenter of the binary
system.  Thus, the systemic velocity must be determined to correct the
radial velocities to the heliocentric system. To do so, we considered the
individual disentangled HST/STIS spectra and followed three different
approaches; namely, gaussian fitting to well-resolved lines,
cross-correlation against an observed spectrum, and cross-correlation
against a synthetic spectrum. For the first approach, line identification
was accomplished through comparison with the spectral atlas of Rogerson
(1985).  When cross-correlating with an observed spectrum, the star chosen
as a template was the SMC star AzV 304 (B0.5~V) which was observed with
HST/STIS in the same spectral region (dataset identifier O51901030). All
three approaches yielded very similar results, with systemic radial
velocities of $v_{\gamma}=$ 257.7, 263.4, and 260.3~km~s$^{-1}$,
respectively. The value finally adopted,
$v_{\gamma}=261.4\pm4.6$~km~s$^{-1}$, is the weighted average of these
three with the corresponding uncertainty.

The final heliocentric radial velocities derived from all the HST/STIS
spectra using the procedure outlined above are listed in Table
\ref{tabRV}, along with the date of observation and the corresponding
phase. The individual errors of the velocities are not provided by {\it
KOREL} and a reliable estimation is not straightforward. This issue is
addressed in more detail in \S \ref{secRV}. As a reality check on our
results, individual radial velocities were also measured independently
using two additional procedures: gaussian fits to the strong Si~{\sc iii}
and C~{\sc ii} features, and cross-correlation of every spectrum against
the HST/STIS spectrum of AzV 304. Both methods yielded velocities
compatible with those in Table \ref{tabRV}, although with significantly
larger uncertainties.

\subsection{UV/Optical Spectrophotometry}

We obtained spectrophotometric observations of EROS~1044 at UV and optical
wavelengths with the Faint Object Spectrograph (FOS) aboard HST on 20
December 1996, at binary phase 0.138. Data were obtained in four
wavelength regions, using the G130H, G190H, G270H, and G400H observing
modes of the FOS with the 3.7\arcsec$\times$1.3\arcsec\ aperture, yielding
a spectral resolution of $\lambda/\Delta\lambda \simeq 1300$. The dataset
identifiers are Y3FU0B03T, Y3FU0B06T, Y3FU0B05T, and Y3FU0B04T,
respectively.  The data were processed and calibrated using the standard
pipeline processing software for the FOS.  As will be discussed further in
\S 3.3, each of the FOS spectra was scaled, by a mean amount of +2.7\%, to
normalize them to a fiducial phase of 0.25. The four individual
observations were then merged to form a single spectrum which covers the
range 1145 \AA\ to 4790 \AA.

Additional HST spectrophotometric observations of EROS~1044 were carried
out with STIS to extend the wavelength coverage. Data were obtained on 01
January 2001 (phase 0.30) and 25 December 2000 (phase 0.80) in the G430L
and G750L observing modes, respectively, with the
52\arcsec$\times$0.5\arcsec\ aperture (yielding a spectral resolution of
$\lambda/\Delta\lambda \simeq 750$). The dataset identifiers are O665A1030
and O665A2030, respectively. The data were processed and calibrated using
the standard pipeline processing software. Cosmic ray blemishes were
cleaned manually and the spectra were trimmed to the regions 3210--5690
\AA\/ and 5290--7490 \AA\/ for the G430L and G750L data, respectively. The
two spectra were scaled, by less than +1\%, to normalize them to phase
0.25.  Because of concerns about photometric stability, the two STIS
spectra were not merged (see \S 3.3).

\section{The Analysis}

Our study of EROS 1044 proceeds from three separate but interdependent
analyses. These involve the light curve, the radial velocity curve, and
the spectral energy distribution (SED). The combined results provide
essentially a complete description of the physical properties of the
EROS~1044 system and an accurate measurement of its distance. Each of the
three analyses is described below.

\subsection{The Light Curve} \label{seclc}

The fits to the light curves were carried out using an improved version of
the Wilson-Devinney (W-D) program (Wilson \& Devinney 1971) that includes
a model atmosphere routine developed by Milone, Stagg \& Kurucz (1992) for
the computation of the stellar radiative parameters, based on the {\it
ATLAS9} code. A detached configuration was chosen for all solutions. Both
detailed reflection model (MREF=2, NREF=1) and proximity effect
corrections were taken into account when fitting the light curves. The
bolometric albedo and the gravity-brightening coefficients were both set
at the canonical value of 1.0 for stars with radiative envelopes. For the
limb darkening we used a logarithmic law as defined in Klinglesmith \&
Sobieski (1970), with first- and second-order coefficients interpolated at
each iteration for the exact $T_{\rm eff}$ and $\log g$ of each component
from a set of tables computed in advance using a grid of {\it ATLAS9}
model atmospheres. A mass ratio of $q=M_{\rm S}/M_{\rm P}=1.178$ was
adopted from the spectroscopic solution (\S \ref{secRV}), and the
temperature of the primary component was set to 21400~K, as discussed in
\S \ref{secSED}. A circular orbit was adopted for the system as suggested
by the equal width of the eclipses and the occurrence of the secondary
minimum at phase 0.5. Test runs of W-D with eccentricity and argument of
the periastron as free parameters yielded a value of $e=0.0009\pm0.0012$
thus confirming an effectively null eccentricity (see \S \ref{secphys} for
further explanations). Finally, both components were assumed to rotate
synchronously with the orbital period.

The iterations with the W-D code were carried out automatically until
convergence, and a solution was defined as the set of parameters for which
the differential corrections suggested by the program were smaller than
the internal probable errors on three consecutive iterations. As a general
rule, several runs with different starting parameters are used to make
realistic estimates of the uncertainties and to test the uniqueness of the
solution.

To run the solutions we focused initially on the OGLE light curve, which
exhibits about half the scatter of the EROS light curves. The relative
weight of the observations was set according to the photometric
uncertainties of the magnitude measurements (i.e. ``noise-dependent''
weights). In our initial runs we solved for the following light curve
parameters: the orbital inclination ($i$), the temperature of the
secondary (${T_{\rm eff}}_{\rm S}$), the gravitational potentials
($\Omega_{\rm P}$ and $\Omega_{\rm S}$), the luminosity of the primary
($L_{\rm P}$), and a phase offset ($\Delta\phi$). Convergence was achieved
rapidly but, when analyzing the residuals, a significant systematic trend
near the quadratures became evident. Subsequent tests revealed that the
systematic deviations from the model light curve were caused by the
presence of an additional source of light in the system. Thus, the
fraction of ``3rd light'' was allowed to vary and convergence was reached
for a contribution of $\sim$12.5\% to the total light of the system (at
phase 0.25). Note that, while usually poorly determined, ``3rd light'' can
be fairly easily constrained in this case due to the out-of-eclipse
variations of the light curve caused by the mild ellipticity of the
secondary component. Examination of the acquisition images for the
HST/STIS observations of EROS1044 (\S 2.2) revealed that the source of the
``3rd light'' is unresolved at a level of $\sim0\farcs2$.
 
Using the solution discussed above as starting point, we ran W-D fits to
all three available light curves simultaneously and allowing for variable
``3rd light'' contributions. The relative weights of the light curves were
set iteratively and adopted as the r.m.s. residuals of the W-D fits. These
turned out to be 0.014, 0.013 and 0.006~mag for the EROS $B_{\rm E}$, EROS
$R_{\rm E}$, and OGLE $I$ light curves, respectively. The resulting
orbital and physical parameters are well-defined and the best-fitting
model light curves, together with the O--C residuals for the OGLE
photometry, are shown in Figure \ref{figLC}.

The final orbital and stellar parameters adopted from the light curve
analysis are listed in Table \ref{tabPARMS}. The uncertainties given in
this table were not adopted from the formal probable errors provided by
the W-D code, but instead from numerical simulations and other
considerations. Several sets of starting parameters were tried in order to
explore the full extent of the parameter space. Special attention was paid
to the error contribution from the uncertainty in the fraction of ``3rd
light''. In addition, the W-D iterations were not stopped after a solution
was found, instead, the program was kept running to test the stability of
the solution and the geometry of the $\chi^2$ function near the minimum.
The scatter in the resulting parameters from numerous additional solutions
yielded estimated uncertainties that we consider to be more realistic, and
are generally several times larger than the internal statistical errors. 

Previous analyses of the EROS $B_{\rm E}$ and $R_{\rm E}$ light curves had
been carried out by Kang et al. (1997) and Tobin et al. (1997) using W-D.  
These authors, however, had no radial velocity data to determine the mass
ratio of the system. Kang et al. ran solutions with $q=1$ and Tobin et al.  
were able to constrain the mass ratio within $q=1\pm0.2$. More
importantly, neither of the two analyses introduced a ``3rd light''
contribution in the fitting procedure. Because of this, the orbital and
stellar properties presented in these works differ significantly from
those of the current paper.

\subsection{The Radial Velocity Curve}\label{secRV}

\begin{figure*}
\plotone{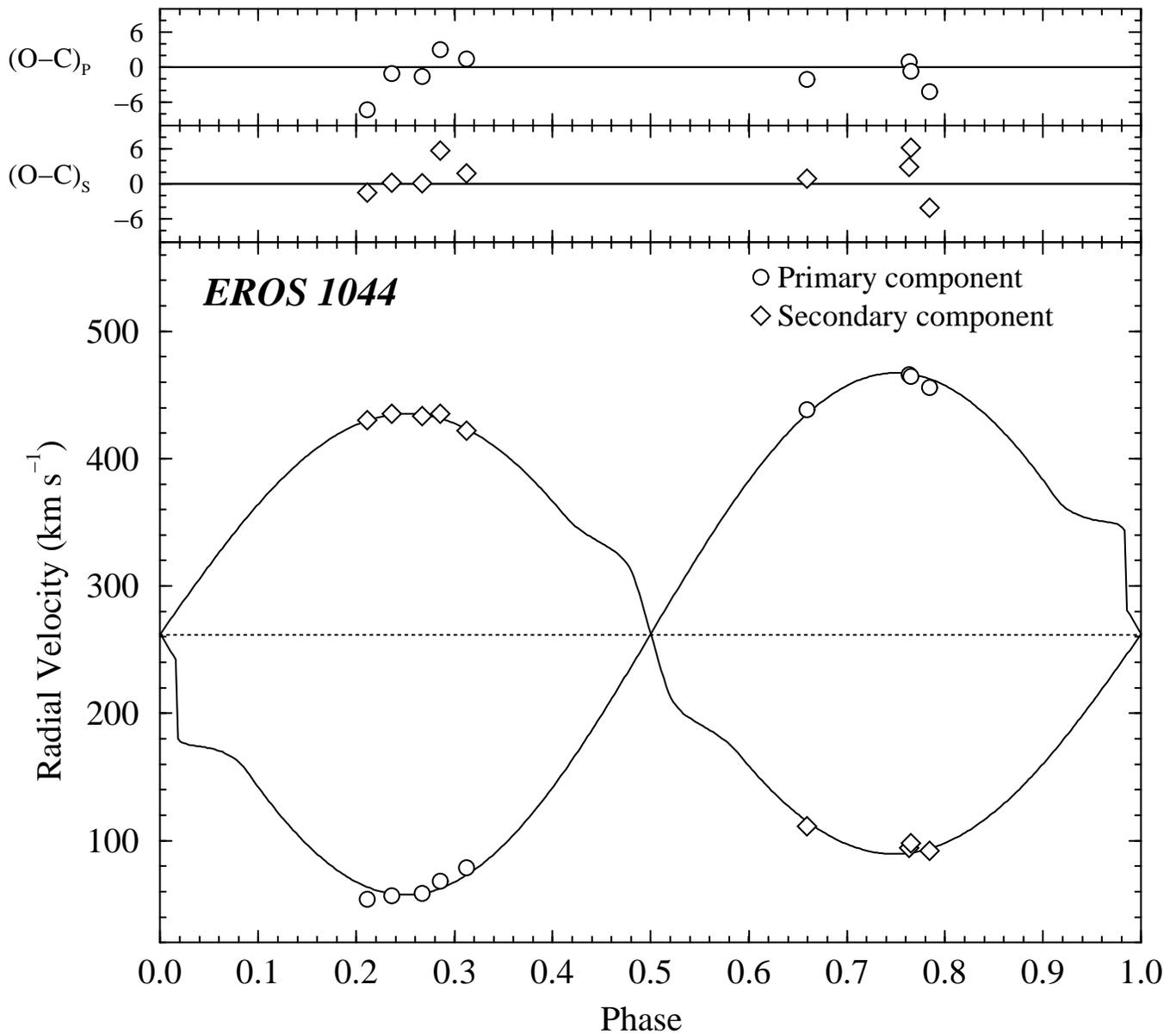}
\caption[f4.eps]{Radial velocity data for EROS~1044 (see Table
\ref{tabRV}) superimposed with best-fitting model. The parameters derived from
the data are listed in Table \ref{tabPARMS}. The fit assumed a circular orbit.
The residuals to the fit are shown above the radial velocity curve and 
indicate r.m.s.  uncertainties in the data of $\sim$3.4 and 
$\sim$3.1~km~s$^{-1}$ for the primary and secondary components, respectively.
\label{figRV}}
\end{figure*}

The radial velocity curve was fit using the same version of the W-D
program as described above. The free parameters were: the orbital
semi-major axis ($a$), the mass ratio ($q\equiv M_{\rm S}/M_{\rm P}$), and
a velocity zero point (the systemic radial velocity $v_{\gamma}$).  (Note
that fitting $a$ and $q$ is equivalent to fitting the velocity
semiamplitudes $K_{\rm P}$ and $K_{\rm S}$.) The rest of the parameters
were set to those resulting from the light curve solutions discussed in
the previous section. The best fit to the radial velocity curve is shown
in Figure \ref{figRV}.  The fit residuals (indicated as ``O--C'' in Figure
\ref{figRV} and in the last two columns of Table \ref{tabRV})  correspond
to r.m.s. errors of 3.1 and 3.3~km~s$^{-1}$ for the primary and secondary
components, respectively.  These small internal errors reflect the
excellent performance of the disentangling technique.

The parameters resulting from the radial velocity curve fit are listed in
Table \ref{tabPARMS}. The uncertainties given in the table are not taken
directly from the W-D output, since they fail to account for any
systematic effects that could be present in the velocity data. Instead, we
estimated more realistic errors by considering the scatter of the
velocities derived from the disentangling analysis of separate spectral
regions. In addition, the STIS Instrument Handbook quotes a 2$\sigma$
uncertainty in the absolute wavelength scale of 0.5--1.0 pix. This means,
in our case, a standard error of about 4~km~s$^{-1}$. By taking all these
considerations into account, the uncertainties of the velocity
semiamplitudes $K_{\rm P}$ and $K_{\rm S}$ were adopted as 5.1 and
5.2~km~s$^{-1}$, respectively, and the errors for the rest of the radial
velocity curve parameters in Table \ref{tabPARMS} were scaled accordingly.

\subsection{The UV/Optical Energy Distribution} \label{secSED}
\subsubsection{The Fitting Procedure}

The observed SED $f_{\lambda\oplus}$ of a binary system depends on the
surface fluxes of the individual components, the stellar radii, and on the
attenuating effects of distance and interstellar extinction. Modeling the
shape and level of the SED over a wide wavelength region can therefore
yield determinations of both intrinsic and extrinsic properties of the
system.

Specifically, the observed flux can be expressed as:
\begin{eqnarray} \label{basic1}
f_{\lambda\oplus} &=&\left(\frac{R_{\rm P}}{d} \right)^2 [F_{\lambda}^{\rm P} + (R_{\rm S}/R_{\rm P})^2 
F_{\lambda}^{\rm S}] \times 10^{-0.4 E(B-V) [k(\lambda-V) + R(V)]}
\end{eqnarray}
where $F_{\lambda}^i$ $\{i={\rm P},{\rm S}\}$ are the surface fluxes of the
primary and secondary stars, the $R_i$ are the absolute radii of the
components, and $d$ is the distance to the binary.  The last term
carries the extinction information, including  $E(B-V)$, the normalized
extinction curve $k(\lambda-V)\equiv E(\lambda-V)/E(B-V)$, and the
ratio of selective-to-total extinction in the $V$ band $R(V) \equiv
A(V)/E(B-V)$.  We represent the stellar surface fluxes with R. L.
Kurucz's {\it {\it ATLAS9}} atmospheres and use a parameterized
representation of UV-through-IR extinction based on the work of
Fitzpatrick \& Massa (1990) and Fitzpatrick (1999; hereafter F99).  The
Kurucz models are each functions of four parameters ($T_{\rm eff}$,
$\log g$, [m/H], and microturbulence velocity $\mu$), and the
extinction curves are functions of six parameters (see F99).

We model the observed SED by performing a non-linear least squares fit to
determine the best-fit values of all parameters which contribute to the
right side of equation 1. For EROS~1044 (as in Papers I and II), we can
make several simplifications which reduce the number of free parameters in
the problem: (1) the temperature ratio of the two stars is known from the
light curve analysis; (2) the surface gravities can be determined by
combining results from the light and radial velocity curve analyses and
are log g = 4.04 and 3.72 for the primary and secondary stars,
respectively (see \S 5); (3) the values of [m/H] and $\mu$ can be assumed
to be identical for both components; (4) the ratio $R_{\rm S}/R_{\rm P}$
is known; and (5) the standard mean value of $R(V) = 3.1$ found for the
Milky Way can reasonably be assumed given the existing LMC measurements
(e.g., Koornneef 1982; Morgan \& Nandy 1982; see \S 4).

In contrast to our previous studies, the two components of EROS~1044 have
significantly different properties, with the secondary star being about
twice as bright as the primary at all observed wavelengths.  As a result,
its properties (particularly its radius) are better-determined than the
primary's. Since equation (\ref{basic1}) is symmetric to the identity of
the primary and secondary stars, we therefore reference our SED analysis
to the secondary star. Given the simplifications noted above, we model
EROS~1044's SED by solving for the best-fitting values of $T_{\rm
eff}^{\rm S}$, [m/H]$_{\rm PS}$, $\mu_{\rm PS}$, $(R_{\rm S}/d)^2$,
$E(B-V)$, and six $k(\lambda-V)$ parameters --- a total of 11 free
parameters.

We prepared the three spectrophotometric datasets (one merged FOS spectrum
and two STIS spectra) for the SED analysis by (1) velocity-shifting to
bring the centroids of the stellar features to rest velocity; (2)
correcting for the presence of a strong interstellar H~{\sc i} Ly$\alpha$
absorption feature in the FOS spectrum at 1215.7~\AA\ (see \S 4); and (3)
binning to match the {\it ATLAS9} wavelength grid. The statistical errors
assigned to each bin were computed from the statistical errors of the
original data, i.e., $\sigma_{bin}^2 = 1/\Sigma(1/\sigma_i^2)$, where the
$\sigma_i$ are the statistical errors of the individual spectrophotometric
data points within a bin. For all the spectra, these binned uncertainties
typically lie in the range 0.5\% to 1.5\% of the binned fluxes.  The
nominal weighting factor for each bin in the least squares procedure is
given by $w_{bin} = 1/\sigma_{bin}^{2}$.  We exclude a number of
individual bins from the fit (i.e., set the weight to zero) for the
reasons discussed by FM99 (mainly due to the presence of interstellar gas
absorption features).

As discussed in Paper II, we do not merge the FOS and STIS data into a
single spectrum, but rather perform the fit on the three binned spectra
simultaneously and independently.  We assume that the FOS fluxes represent
the ``true'' flux level and account for zero point uncertainties in the
STIS data by incorporating two zeropoint corrections (one for each STIS
spectrum) in the fitting procedure.  We later explicitly determine the
uncertainties in the results introduced by zeropoint uncertainties in FOS.

\subsubsection{Special Considerations for EROS~1044}

The fact that the EROS~1044 light curve is not flat between eclipses ---
resulting from a mild distortion of the secondary star --- complicates the
SED analysis.  Unless all the spectrophotometric data are obtained at the
same phase, brightness offsets of up to $\sim$3\% may occur between
spectra. To correct for this, we adjust all spectra to the phase of
maximum brightness (phase = 0.25), i.e., the point at which each star
displays its maximum cross-section. This correction amounts to the simple
application of a multiplicative scale factor\footnote{The application of a
multiplicative scale factor is granted by the small variations in the
relative light ratio (primary, secondary, and ``3rd light'') between the
phases of the FOS observations and phase 0.25. The relative contribution
of the ``3rd light'' varies by only 0.3\% and changes in the light ratio
between the primary and the secondary amount to a mere 0.1\% effect.},
derived from the $B$, $R$, and $I$ light curves.  The offsets applied to
the HST/FOS and HST/STIS data are: +3.2\% for FOS/G130H, +2.3\% for
FOS/G190H, +2.5\% for FOS/G270H, +2.7\% for FOS/G400H, +0.3\% for
STIS/G430L, and +0.7\% for STIS/750L. The offsets were applied to the FOS
data prior to merging them into a single spectrum.

Because the ``3rd light'' source in the EROS~1044 light curve is
unresolved in the HST acquisition images (see \S 3.1), the flux
contribution is also present in the spectrophotometric data and must be
accounted for in the SED analysis.  The light curve data provide strong
constraints on the nature of this third body. Anticipating the reddening
derived below, the object's fractional contribution at $V$ ($\sim$12.5\%)
implies $V_0 \mbox{(3rd light)} \simeq 17.3$. The constancy of the
fraction of ``3rd light'' from the $B$ through the $I$ bands indicates
that it is an early-B star (like the EROS~1044 system).  More
specifically, $T_{\rm eff}$ must be greater than $\sim$12000 K, or else
the $I$ band would show significantly greater contribution than $B$, and
must be less than $\sim$22000 K, since a hotter star --- even if it were
on the ZAMS --- would be brighter than observed.

With these constraints in mind, we include a constant contribution in the
fit to EROS~1044's SED, produced by an {\it ATLAS9} model with $T_{\rm
eff} = 17000$ K and $\log g = 4.1$, appropriately scaled to produce 12.5\%
of the observed light in the $B$, $R$, and $I$ bands at phase 0.25.  The
adopted $T_{\rm eff}$ is simply the center of the allowed range and the
$\log g$ corresponds to a star of the appropriate brightness (as estimated
from stellar evolution calculations).  The addition of the ``3rd light''
to the SED analysis adds no new free parameters, but does add some
uncertainty to the results. To evaluate this uncertainty, we assume
1$\sigma$ ranges in $T_{\rm eff}$ and flux level of 14000--20000 K and
10--15\%, respectively (i.e., $\pm3000$ K and $\pm$2.5\% around the
nominal values).  The value of $\log g$ has no influence on the results.  
In \S 5 we discuss briefly the possible nature of this star.

\subsubsection{The Results}

As in our previous papers, we computed the final fit to EROS~1044's SED
after adjusting the weights in the fitting procedure to yield a final
value of $\chi^2 = 1$ --- since the statistical errors of the data
under-represent the total uncertainties (see the discussion in Paper II).  
This was accomplished by quadratically adding an uncertainty equivalent to
1.75\% of the local binned flux to the statistical uncertainty of each
flux point.  (Essentially identical results occur if the statistical
errors are simply scaled by a factor of 2.2 to yield $\chi^2 = 1$.)  This
value of 1.75$\%$ gives an indication of the general quality of the fit to
EROS~1044's SED, excluding the effect of statistical noise.  It is
comparable the quality level we have seen in the previous analyses
(1.5$\%$ for HV~2274; 2.0$\%$ for HV~982).

\begin{figure*}
\plotone{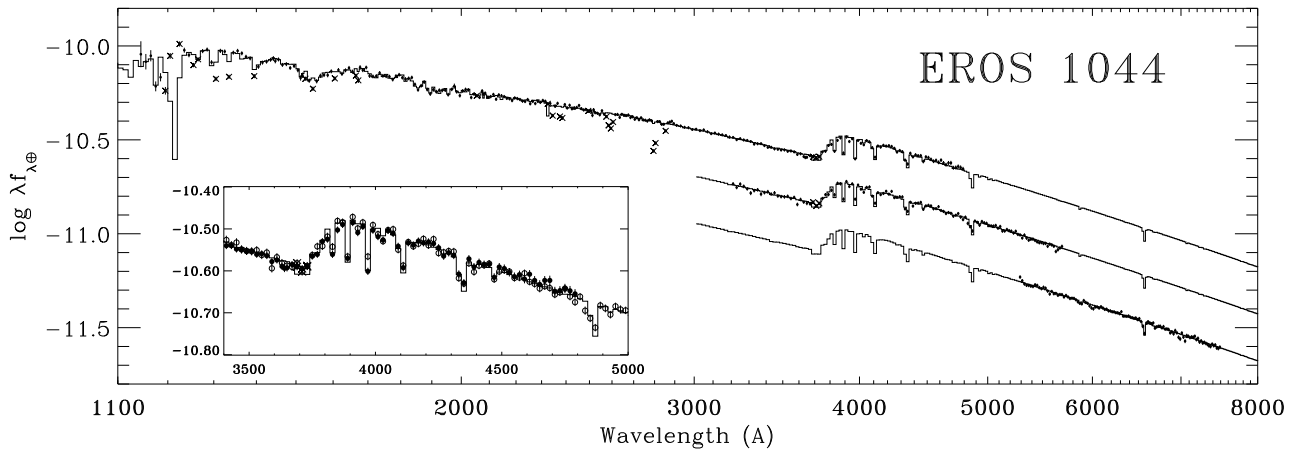}
\caption[f5.eps]{The observed UV/optical energy distribution
of the EROS~1044 system (small filled circles) superimposed with the
best-fitting model. The top spectrum shows the  FOS data, the middle spectrum
(shifted by $-0.25$ dex) the STIS/G430L data, and the lower spectrum (shifted
by $-0.5$ dex) the STIS/G750L data. Vertical lines through the data points
indicate the $1\sigma$ observational errors. The energy distribution fitting
procedure was performed simultaneously on all three datasets.  Crosses denote
data points excluded from the fit, primarily due to contamination by
interstellar absorption lines.  The various constraints imposed on the fit are
discussed in \S 3.3. The model, shifted to match the spectra, consists of a
trio of reddened and distance-attenuated Kurucz {\it ATLAS9} atmosphere models
(histogram-style lines), corresponding to the primary and secondary components
of the binary system and the ``3rd light'' component.  The parameters derived
from the fit to the energy distribution are listed in Tables \ref{tabPARMS},
\ref{tabSTIS}, and \ref{tabEXT}.  The inset shows a blowup of the region
surrounding the Balmer Jump which illustrates the overlap between the FOS
(solid circles) and STIS/G430L (open circles) data. \notetoeditor{THIS FIGURE
IS INTENDED TO SPAN TWO COLUMNS}
\label{figSED}}
\end{figure*}

The best-fitting values of the energy distribution parameters for
EROS~1044 and their 1-$\sigma$ uncertainties (``internal errors'') are
listed in Tables \ref{tabPARMS} (stellar properties), \ref{tabSTIS} (STIS
offsets), and \ref{tabEXT} (extinction curve parameters).  The internal
errors were evaluated by quadratically combining (for each parameter) the
uncertainties from the fitting procedure (computed using the full
correlations) with the errors induced by variations of the $T_{\rm eff}$
and flux level of the ``3rd light'' within the 1-$\sigma$ ranges specified
above. A comparison between the observed spectra and the best-fitting
model is shown in Figure \ref{figSED}.  The three binned spectra are
plotted separately in the figure for clarity (small filled circles).  The
zeropoint offset corrections (see Table \ref{tabSTIS}) were applied to all
STIS spectra in Figure \ref{figSED}.  Note that we show the quantity
$\lambda$f$_{\lambda\oplus}$ as the ordinate in Figure \ref{figSED}
(rather than f$_{\lambda\oplus}$) strictly for display purposes, to
``flatten out'' the energy distributions.

The correction factors of 6.7\% and 4.5\% required to rectify the STIS
G430L and G750L spectra, respectively, are similar to the results found
for HV~982 in Paper II.  This apparently systematic effect probably
results from light-loss in the STIS $0\farcs5$ slit.  This will be tested
by using a wider slit in future STIS observations.

\section{The Interstellar Medium Towards EROS~1044}

We measure the column density of interstellar H~{\sc i} in the foregrounds
of the stars in our program by comparing their observed H~{\sc i}
Ly$\alpha$ 1216.7 \AA\/ absorption line profiles with theoretical profiles
consisting of a synthetic stellar spectrum convolved with an interstellar
absorption profile. The interstellar profile is constructed assuming a
component at 0 km s$^{-1}$ with $N$(H~{\sc i}) $=
5.5\times10^{20}$~cm$^{-2}$, corresponding to Milky Way foreground gas
(see, e.g., Schwering \& Israel 1991), and a second component with a
LMC-like velocity and a column density which is varied to produce the best
fit to the data (as judged by visual inspection).

\begin{figure*}
\plotone{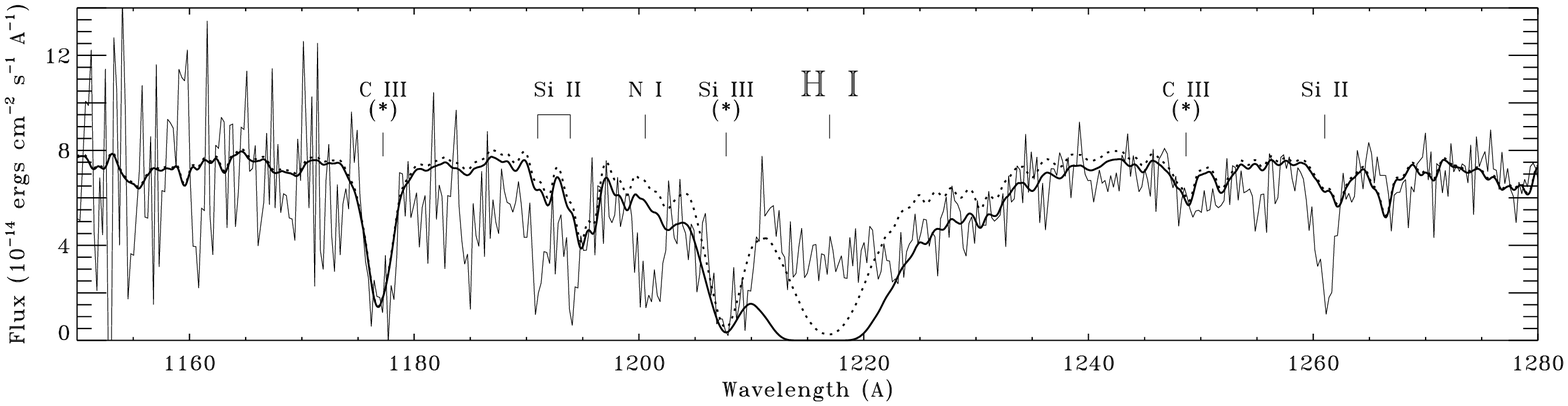}
\caption[f6.eps]{Derivation of the interstellar H~{\sc i} column
density towards EROS~1044.  The FOS data centered on the H~{\sc i} Ly$\alpha$
line at 1215.7 \AA\/ are shown (thin solid line).  Prominent stellar
features (denoted with an asterisk) and interstellar features are
labeled.  The dotted line represents a synthetic spectrum of the EROS~1044
system, constructed by combining two individual velocity-shifted
spectra. (The ``3rd light'' contribution is negligible in this region.) 
The individual spectra were computed using Ivan Hubeny's {\it
SYSNPEC} spectral synthesis program with Kurucz {\it ATLAS9} atmosphere
models of the appropriate stellar parameters as inputs.  The solid
curve shows the synthetic spectrum convolved with an interstellar H~I
Ly$\alpha$ line computed with a Galactic foreground component of ${\rm
N(H~I) = 5.5\times10^{20} cm^{-2}}$ at 0 km s$^{-1}$ and a LMC
component of ${\rm N(H~I) = 2.0\times10^{20} cm^{-2}}$ at 257 km
s$^{-1}$ (see text in \S 4).
\label{figHI}}
\end{figure*}

The best-fitting Ly$\alpha$ profile for EROS~1044 can be seen in Figure
\ref{figHI}, where we illustrate the unbinned FOS spectrum, the synthetic
stellar spectrum (dotted line) and the convolution of the synthetic
spectrum with the interstellar profile (thick solid curve). The LMC
component of the interstellar profile is located at ${\rm v} =
257$~km~s$^{-1}$ (from the H~{\sc i} 21 cm results of Rohlfs et al. 1984)
with a best-fitting column density of $2.0\times10^{20}$~cm$^{-2}$.  The
relatively noisy FOS spectrum, strong Ly$\alpha$ geocoronal contamination
(which fills in the bottom of the line), and the strength of the stellar
feature conspire to make this measurement somewhat uncertain. We estimate
a $\sim2\sigma$ range of $\sim\pm2\times10^{20}$~cm$^{-2}$, i.e., it is
barely possible that EROS~1044 is ``above'' all the LMC H~{\sc i} in its
direction (although this is not consistent with the presence of LMC
reddening, as discussed below), and unlikely that it is behind more than
$\sim4\times10^{20}$~cm$^{-2}$ of LMC H~{\sc i}.

The value of ${\rm E(B-V)}$ found for EROS~1044 in \S 3, combined with a
Galactic foreground reddening of $E(B-V)_{\rm MW}=0.055$ (from the data of
Oestreicher, Gochermann, \& Schmidt-Kaler 1995), indicates a LMC reddening
corresponding to $E(B-V)_{\rm LMC}=0.011$. The implied LMC gas-to-dust
ratio of $1.8\times10^{22}$~cm$^{-2}$~mag$^{-1}$ --- with a large
uncertainty --- is compatible with other measurements (see Fitzpatrick
1985).

The H~{\sc i} 21-cm emission line survey of Rohlfs et al. (1984) reveals
that the {\it total} LMC H~{\sc i} column density along the EROS~1044 line
of sight is about $1.5\times 10^{21}$~cm$^{-2}$, centered at
257~km~s$^{-1}$ (as noted above).  This result, plus the Ly$\alpha$
absorption and $E(B-V)_{\rm LMC}$ determinations, indicates that EROS~1044
is embedded within (as opposed to being behind or in front of) the main
H~{\sc i} mass along its line of sight.  In addition, the systemic
velocity for EROS~1044, 261~km~s$^{-1}$, is nearly identical to that of
the H~{\sc i} gas.  Thus, like HV2274 (which is also located in the LMC's
Bar) --- and unlike HV~982 --- it is clear that EROS~1044 is physically
associated with the LMC's gas and dust disk.
 
\begin{figure*}
\plotone{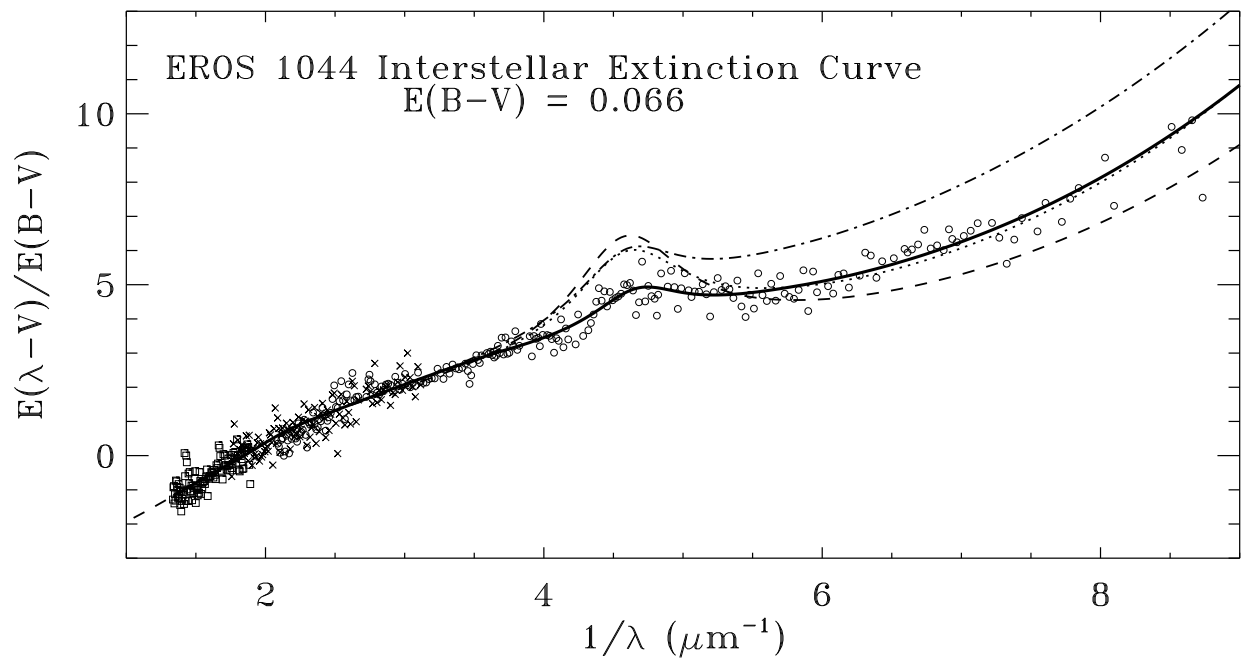}
\caption[f7.eps]{Normalized UV-through-optical interstellar
extinction curve for EROS~1044.  The thick solid line shows the
parameterized form of the extinction curve as determined by the SED
fitting procedure.  The recipe for constructing such a ``custom''
extinction curve is taken from F99 and the parameters defining it are
listed in Table \ref{tabEXT}.  Small symbols indicate the actual
normalized ratio of model fluxes to observed fluxes: circles, crosses,
and squares indicate FOS, STIS G430L, and STIS G750L data,
respectively.  Shown for comparison are the mean Milky Way extinction
curve for R = 3.1 from F99 (dashed line) and the mean LMC and 30
Doradus curves from Fitzpatrick (1986; dotted and dash-dotted lines,
respectively).  The EROS~1044 curve arises from dust in both the Milky
Way (${\rm E(B-V)_{MW} \simeq 0.055}$) and the LMC (${\rm E(B-V)_{LMC}
\simeq 0.011}$).  The main attributes of the curve, its high far-UV
level and extremely weak 2175 \AA\/ bump, are likely shared
characteristics of the extinction in both environments.
\label{figEXT}}
\end{figure*}

The interstellar extinction curve determined for the EROS~1044 sightline
is shown in Figure \ref{figEXT}.  Small symbols indicate the normalized
ratio of model fluxes to observed fluxes, while the thick solid line shows
the parameterized representation of the extinction, which was actually
determined by the fitting process. The parameters defining the curve are
listed in Table \ref{tabEXT}. The $E(B-V)$ results above indicate that
this curve is dominated by Milky Way foreground dust, but contains a
$\sim$20\% contribution from LMC dust. The curve is nearly identical to
the result for HV~982 in Paper II, which is also dominated by Milky Way
foreground dust, and notable for the very weak 2175 \AA\/ bump. We have
been granted FUSE observations for EROS~1044 that will allow us to extend
the analysis into the far-UV. We will discuss the interstellar extinction
results for our whole sample of LMC EBs in a future paper.

\section{The Physical Properties of the EROS 1044 Stars} \label{secphys}

The results of the analyses described above can be combined to provide a
detailed characterization of the physical properties of the EROS 1044
system. We summarize these properties in Table \ref{tabSTAR} and indicate
in the Notes to the Table how the individual stellar properties were
derived.

It is important to note that the results presented so far have been
obtained independently of theoretical evolution considerations. Thus,
stellar evolution models can be used to provide a valuable check on the
self-consistency of our empirical results. To make this comparison, we
utilize the evolutionary models of Claret (1995, 1997) and Claret \&
Gim\'enez (1995, 1998), together referred to as the CG models.  These
models cover a wide range in both metallicity ($Z$) and initial helium
abundance ($Y$), incorporate the most modern input physics, and adopt a
convective overshooting parameter of 0.2~H$_{\rm p}$.

\begin{figure*}
\plotone{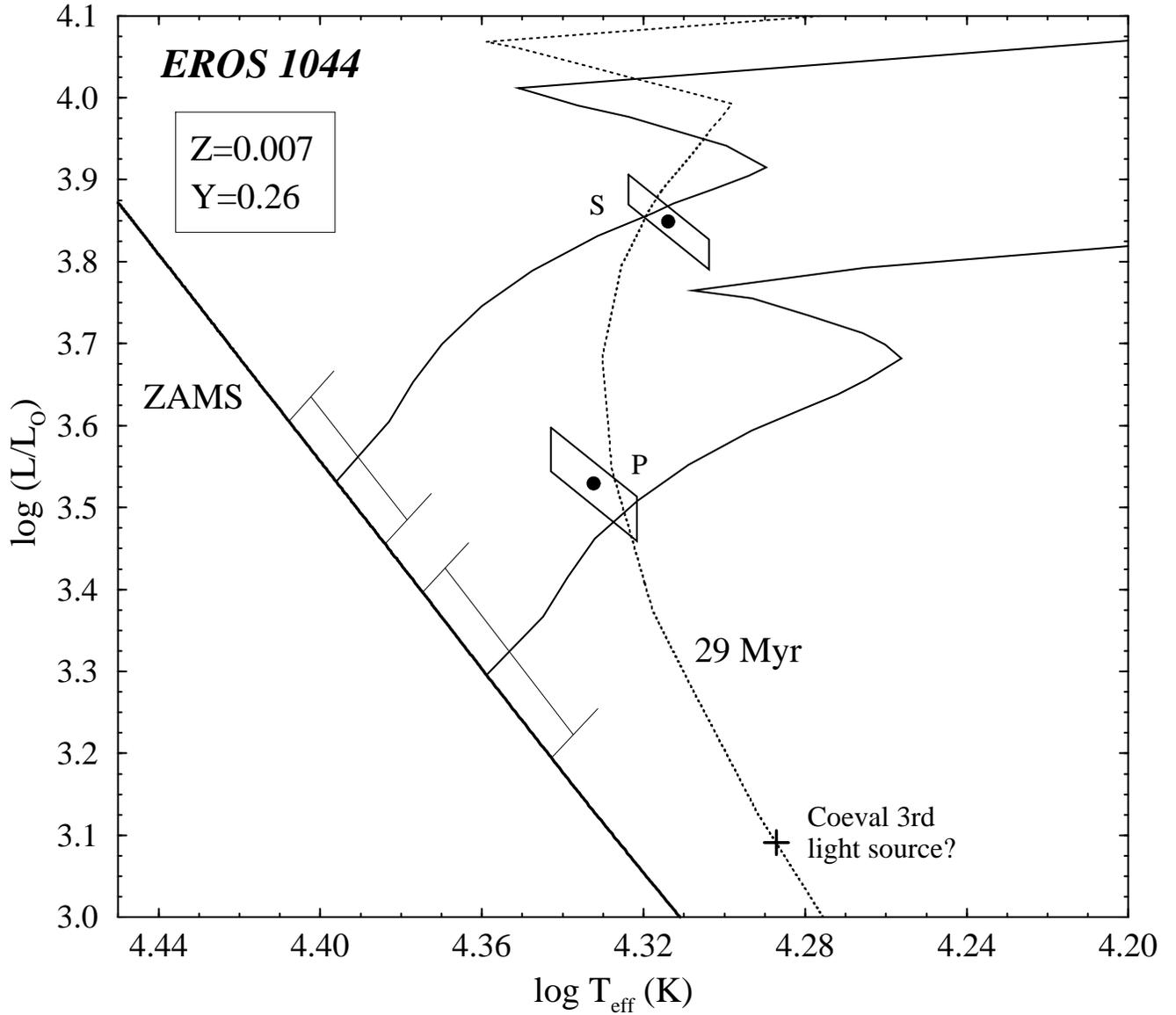}
\caption[f8.eps]{A comparison of the EROS~1044 results with
stellar evolution theory. The positions of the the primary (P) and
secondary (S) components on the $\log L$ vs. $\log T_{\rm eff}$ diagram
are indicated by the filled circles, with the skewed rectangles
representing the 1$\sigma$ error boxes. The two stellar evolution
tracks shown (solid curves) correspond to the masses derived from the
binary analysis. Error bars on the tracks near the ZAMS show the effect
of $1\sigma$ uncertainties in the masses.  The dotted line shows an
isochrone corresponding to an age of 29 million years. The source and
properties of the evolution tracks are discussed in \S 5.  The plus sign
indicates the position the ``3rd light'' component would occupy if it were
coeval with the EB system.
\label{figHRD}}
\end{figure*}

The locations of the EROS~1044 components in the $\log T_{\rm eff}$ vs.
$\log L$ diagram are shown in Figure \ref{figHRD}.  The skewed rectangular
boxes indicate the 1$\sigma$ error locus (recall that errors in $T_{\rm
eff}$ and $L$ are correlated).  The best-fitting pair of evolution tracks
are also shown on the plot.  These tracks correspond to the masses derived
from the analysis (see Table \ref{tabSTAR}).  Approximate error bars for
the evolution tracks, which reflect the 1$\sigma$ uncertainties on the
observed masses, are shown near the ZAMS.  Only the helium ($Y$) and metal
abundances ($Z$) were adjusted to optimize the fit.  Best-fitting values
were obtained through the procedure described by Ribas et al. (2000b) and
standard errors are estimated through a Monte Carlo simulation.  The
procedure yielded values of $Y=0.26\pm0.03$ and $Z=0.007\pm0.003$ (i.e.
$\mbox{[m/H]}=-0.40^{+0.16}_{-0.24}$).

The results in Figure \ref{figHRD} demonstrate excellent agreement between
our derivation of the EROS~1044 stellar properties and expectations from
stellar evolution.  The stars' positions in the $\log T_{\rm eff}$ vs.
$\log L$ diagram are consistent with their measured masses; the
metallicity derived from the evolution tracks agrees with that obtained
from the spectrophotometric analysis (see \S \ref{secSED}); and the helium
abundance agrees with expectations from empirical chemical enrichment laws
(see Ribas et al.  2000b).  Further, the stars are consistent with a
single isochrone, indicating an age for the system of $29\pm4$ Myr (dotted
line in Fig. \ref{figHRD}).  We conclude that the EROS~1044 EB system
consists of a pair of normal early-B stars, whose properties are
well-modeled by current stellar atmosphere and structure theory.

There are compelling observational evidences indicating that the orbit of
EROS~1044 is circular (see \S \ref{seclc}). From the data described above
this can be further checked by performing theoretical tidal evolution
calculations. Using the method described in Claret, Gim\'enez, \& Cunha
(1995) we find that the system components synchronized their rotational
velocities very soon ($\sim$1 Myr) after the ZAMS. Additionally, our
calculations indicate that the orbit became circular at an age of 17~Myr
(compared with the current 29~Myr) thus confirming the observational
evidences.

Numerous EB systems have been found to be part of hierarchical triple
systems, with companions in more distant orbits.  Thus, it is reasonable
to suspect that the ``3rd light'' contributing to the EROS~1044 light
curve might arise from a coeval companion physically associated with the
EB system.  Based on the 29 Myr isochrone and on the observed brightness
of the ``3rd light'' component, the properties of such a companion would
be:  $M=5.7$~M$_{\sun}$, $T_{\rm eff} = 19400$ K, $\log L/{\rm L}_{\sun} =
3.09$, and $\log g = 4.20$.  The position of this star on the $\log T_{\rm
eff}$ vs.  $\log L$ diagram is indicated in Figure \ref{figHRD} by the
plus sign.  The temperature of the putative coeval companion is within the
observationally-allowable range (see \S 3.3), and thus it is plausible ---
although not proven --- that EROS~1044 is part of a triple system. Further
insight into the nature of the 3rd body might be derived from the analysis
of spectrophotometry carried out during the total occultation at 0.0
phase, when the light contribution of one of the eclipsing components is
not present. Also, additional determinations of eclipse times in the
future could reveal the presence (and yield the mass) of the third body
through the analysis the light travel time effect arising from the
movement of the eclipsing pair around the barycenter of the triple system.

\section{The Distance to EROS~1044 and the LMC}

The discussion above in \S 5 demonstrates that the properties of the
EROS~1044 EB system are very well-characterized, are consistent with a
variety of internal and external reality checks, and reveal the system to
consist of a pair of normal, mildly evolved early B-type stars.  For such
a system, the determination of the distance is straightforward: we combine
the absolute radius of the secondary star $R_{\rm S}$ (derived from the
classical EB analysis) with the distance attenuation factor $(R_{\rm
S}/d)^2$ (derived from the SED analysis) and find $d_{\rm EROS~1044} =
47.5\pm1.8$ kpc, corresponding to a distance modulus of $(V_0 - M_V)_{\rm
EROS~1044} = 18.38\pm0.08$.

The uncertainty in the EROS~1044 distance is estimated from considering
three sources of error: (1) the internal measurement errors in $R_{\rm S}$
and $(R_{\rm S}/d)^2$ given in Table \ref{tabPARMS}; (2) uncertainty in
the appropriate value of the extinction parameter $R(V)$; and (3)
uncertainty in the FOS flux scale zeropoint due to calibration errors and
instrument stability.  Straightforward propagation of errors shows that
these three factors yield individual uncertainties of $\pm 1.60$ kpc, $\pm
0.40$ kpc (assuming $\sigma R(V) = \pm0.3$), and $\pm 0.60$ kpc (assuming
$\sigma f(FOS) = \pm2.5$\%), respectively.  The overall 1$\sigma$
uncertainty quoted above is the quadratic sum of these three errors, which
are considered to be independent. Recall that the internal errors for
$(R_{\rm S}/d)^2$ listed in Table \ref{tabPARMS} already include a
conservative estimate of the uncertainty introduced by the presence of
``3rd light'' in the system.  The only ``adjustable'' factor in the
analysis is the extinction parameter $R(V)$, for which we have assumed the
value 3.1.  The weak dependence of our result on this parameter is given
by: $(V_0 - M_V)_{\rm EROS~1044} = 18.38 - 0.055 \times [R(V)-3.1]$.

In Table \ref{tabDIST} we summarize the distances derived for the three
LMC EB systems we have analyzed so far.  While the distances to the three
EB's are all mutually consistent --- at a level better than
$\sim1.5\sigma$ --- the results certainly do suggest that HV~982 is the
most distant system, perhaps by as much as several kpc.  This inference is
consistent with the analyses of interstellar H~I towards the stars:
EROS~1044 and HV~2274 have been found to be embedded within the H~I disk
along their lines of sight, while HV~982 is located at an undetermined
distance behind the H~I disk (i.e., its H~I emission and absorption column
densities are identical).

\begin{figure*}
\plotone{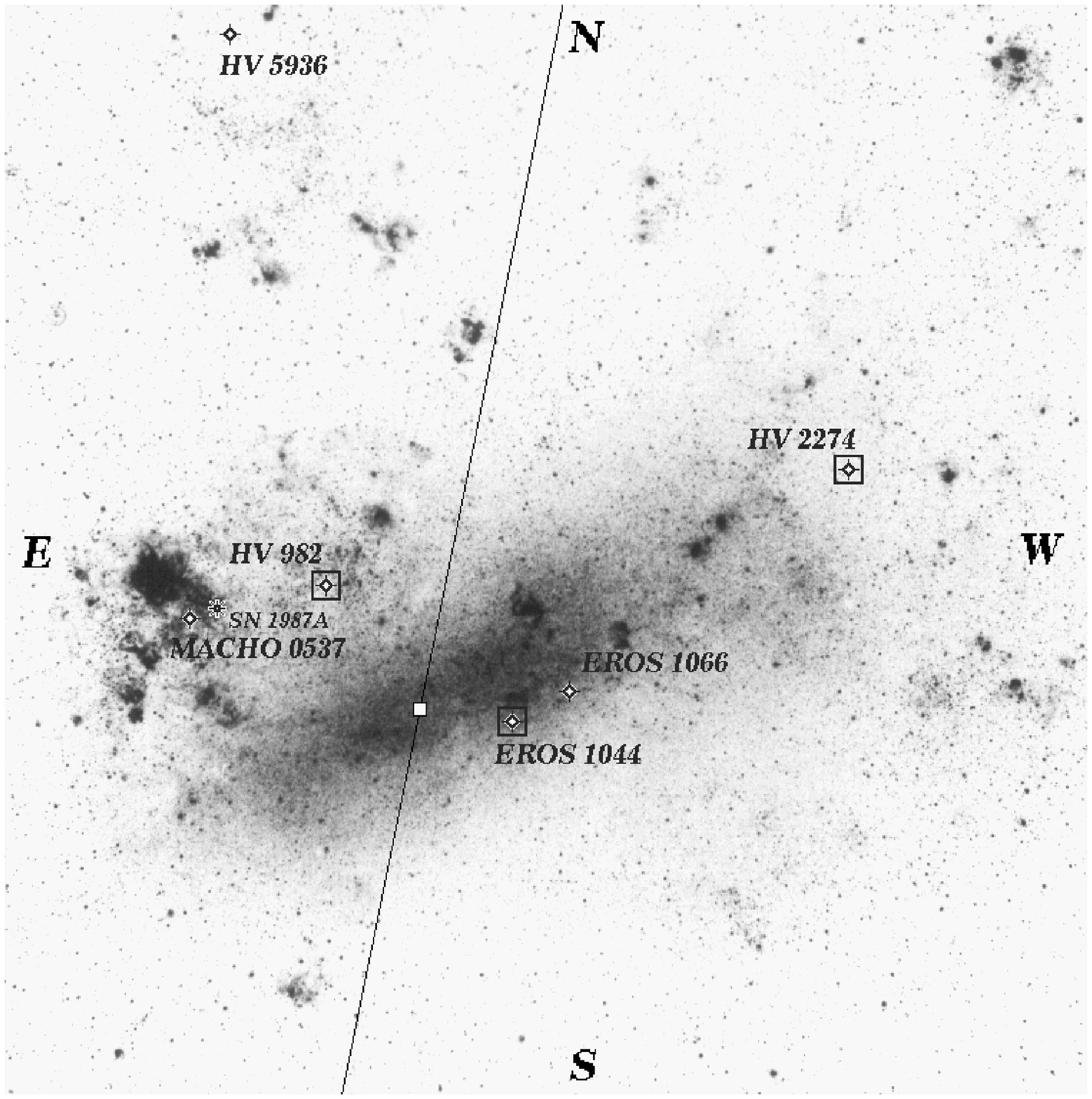}
\caption[f1.eps]{A photo of the Large Magellanic Cloud indicating
the locations of EROS~1044 (this paper), HV~2274 (Paper I), HV~982 (Paper II)
and three targets of future analyses, HV~5936, EROS~1066, and MACHO
053648.7-691700 (labeled in the figure as MACHO 0537). The optical center of
the LMC's Bar according to Isserstedt 1975 is indicated by the open box and the
LMC's line of nodes, according to Schmidt-Kaler \& Gochermann 1992, is shown by
the solid line. The ``nearside'' of the LMC is to the east of the line of
nodes.  The location of SN 1987A is also indicated. Photo reproduced by
permission of the Carnegie Institution of Washington.  \notetoeditor{THIS
FIGURE IS INTENDED TO SPAN TWO COLUMNS}
\label{figLMC}}
\end{figure*}

Since the LMC is not observed face-on in the sky, determining its distance
from those of the individual EB systems requires first choosing a
reference point within the LMC and then correcting the individual results
for the spatial orientation of the LMC. We take the optical center of the
LMC's Bar as our reference point, located at ($\alpha$, $\delta$)$_{1950}$
= ($5^h 24^m$, $-69\arcdeg\/ 47\arcmin$) according to Isserstedt (1975)
and assume a disk-like configuration. For the orientation, we examine two
cases.  First, we assume an inclination angle of $38\arcdeg$ and a
line-of-nodes position angle of $168\arcdeg$ (Schmidt-Kaler \& Gochermann
1992).  (The reference point and this orientation of the line-of-nodes are
indicated in Figure \ref{figLMC} by the open box and solid line,
respectively.  The ``near-side'' of the LMC is eastward of the
line-of-nodes.)  The distances implied for the LMC reference point based
on this assumption are given in column 4 of Table \ref{tabDIST} (``Case
I'').  The second orientation we examine is a recent result from van der
Marel \& Cioni (2001), who found an inclination of $34.7\arcdeg$ and a
line-of-nodes position angle of $122.5\arcdeg$.  This orientation has the
line of nodes lying nearly along the LMC's Bar.  Results based on this
assumption are in column 5 of Table \ref{tabDIST} (``Case II'').

Table \ref{tabDIST} shows that an LMC distance based on HV~2274 is
sensitive to the assumed LMC orientation, while results based on EROS~1044
and HV~982 are largely independent of the assumptions, due to their
apparent proximity to the reference point. It appears that Case II best
preserves the relative agreement among the distances although, given the
size of the uncertainties for EROS~1044 and HV~2274, this is not yet a
rigorous test of the possible LMC orientations.
 
We conclude that our results are thus far compatible with a single
distance of $\sim48$ kpc, corresponding to a distance modulus of
$\sim18.4$ mag.  However, the result for HV~982 provides a tantalizing
hint that there may be some spatial extension of the LMC along its
sightline, which passes near to that of SN~1987A (see Fig. \ref{figLMC}).  
Panagia (1999) finds a distance of $51.4\pm1.2$ kpc for SN~1987A,
consistent with our value for HV~982.  Only further observations will
reveal whether the apparently well-determined HV~982 and SN~1987A
distances are representative of the LMC as a whole or only of a distinct
region (the 30 Doradus complex?), located behind the main mass of the LMC.  
The shorter distances for EROS~1044 and HV~2274 suggest this as a possible
scenario although, as with the discussion of the LMC orientation above,
their large uncertainties preclude firm conclusions.  The precision of the
EROS~1044 result is limited by the uncertain contribution of the ``3rd
light'' component, and is unlikely to be improved.  The HV~2274 result,
however, is limited only by the absence of spectrophotometric data in the
optical spectral region (see Paper II).  The acquisition of such data
could result in an improvement of HV~2274's distance measurement to a
precision comparable to that for HV~982.  While conclusions drawn from
HV~2274 will always be subject to assumptions on the LMC's orientation,
nevertheless, its distance does provide an upper limit to the distance of
the LMC reference point, and could yield a critical counterpoint to HV~982
and SN1987A.
 
In general, our EB analysis is ideally suited to addressing the issues of
the LMC's spatial orientation and internal structure, since it yields
precise results for individual objects which are widely distributed across
the face of the LMC.  We are currently analyzing data for three more
systems, HV~5936, EROS~1066, and MACHO~053648.7-691700 (see Fig.  
\ref{figLMC}).  The apparent locations of EROS~1066 and
MACHO~053648.7-691700 in the Bar and the 30 Doradus region, respectively,
make them ideal for further testing the possibility that stars in the
apparent vicinity of 30 Doradus may lie significantly behind the Bar.  
Within the next few years we hope to expand the program to include about
20 systems.  Recent $n$-body simulations of the tidal interaction of the
Milky Way with the LMC indicate that its structure may be more extended
and complex than presently assumed (see Weinberg 2000). Our ensemble of
targets should be able both to nail down the distance to the LMC and to
provide a detailed probe of its structure and orientation.

\acknowledgements

We thank Villanova undergraduate students J. F. Sepinsky and J. E. Castora
for help with data preparation and analysis. These students were supported
by the Undergraduate Summer Research Assistance Grant from the Delaware
Space Grant College Consortium. Our HST Program Coordinator, Alison Vick,
is warmly thanked for ther invaluable help with the preparation of the
observations. E. L. F. acknowledges support from NASA ADP grant NAG5-7117
to Villanova University and thanks Michael Oestreicher for kindly making
his LMC foreground extinction data available. I. R. acknowledges the
Catalan Regional Government (CIRIT) for financial support through a
postdoctoral Fulbright fellowship. This work was supported by NASA grants
NAG5-7113, HST GO-06683, HST GO-08691, and NSF/RUI AST-9315365.



\begin{deluxetable}{rrrrrr}
\tablewidth{0pc}
\tablecaption{Heliocentric Radial Velocity Measurements for EROS~1044}
\tablehead{
\colhead{HJD}               &
\colhead{Orbital}           &
\colhead{RV$_{\rm P}$}      &
\colhead{RV$_{\rm S}$}      &
\colhead{(O$-$C)$_{\rm P}$} &
\colhead{(O$-$C)$_{\rm S}$} \\
\colhead{($-$2400000)}  &
\colhead{Phase}         &
\colhead{(km s$^{-1}$)} &
\colhead{(km s$^{-1}$)} &
\colhead{(km s$^{-1}$)} &
\colhead{(km s$^{-1}$)} }
\startdata
51899.9747& 0.3189& 78.8&421.7&   1.3 &   1.7 \\
51900.9220& 0.6663&438.5&111.1&$-$2.2 &   0.8 \\
51903.9330& 0.7703&466.1& 94.4&   0.8 &   2.9 \\
51903.9895& 0.7911&455.6& 92.1&$-$4.3 &$-$4.8 \\
51906.6639& 0.7717&464.3& 98.0&$-$0.7 &   6.2 \\
51910.6085& 0.2182& 54.0&430.3&$-$7.3 &$-$1.5 \\
51910.6754& 0.2427& 56.6&435.2&$-$1.1 &   0.2 \\
51910.7601& 0.2737& 58.6&433.5&$-$1.6 &   0.1 \\
51910.8105& 0.2922& 68.3&435.1&   2.9 &   5.6 \\
\enddata
\label{tabRV}
\end{deluxetable}


\small
\begin{deluxetable}{lc}
\tablewidth{0pc}
\tablecaption{Results From Light Curve, Radial Velocity Curve, and 
Spectrophotometry Analyses of EROS~1044}
\tablehead{\colhead{Parameter} & \colhead{Value}}
\startdata
\multicolumn{2}{c}{{\it Light Curve Analysis}}                            \\
Period                                           & $2.727125\pm0.000004$ days\\
Eccentricity                                     & $0.0$ (fixed)          \\
Inclination                                      & $87.2\pm0.9$\phn deg   \\
$T_{\rm eff}^{\rm S}/T_{\rm eff}^{\rm P}$        & $0.958\pm0.008$        \\
$[\mbox{3rd light}]_{{\rm OGLE}~I}$\tablenotemark{a}& $12.6\pm2.5$\%      \\
$[\mbox{3rd light}]_{{\rm EROS}~B}$\tablenotemark{a}& $12.7\pm2.5$\%      \\
$[\mbox{3rd light}]_{{\rm EROS}~R}$\tablenotemark{a}& $12.4\pm2.5$\%      \\
$[L_{\rm S}/L_{\rm P}]_{{\rm OGLE}~I}$           & $2.37\pm0.03$          \\
$[L_{\rm S}/L_{\rm P}]_{{\rm EROS}~B}$           & $2.33\pm0.07$          \\
$[L_{\rm S}/L_{\rm P}]_{{\rm EROS}~R}$           & $2.35\pm0.06$          \\
$r_{\rm P}$\tablenotemark{b}                     & $0.205\pm0.005$        \\
$r_{\rm S}$\tablenotemark{b}                     & $0.323\pm0.003$        \\
$r_{\rm S}/r_{\rm P}$\tablenotemark{c}           & $1.575\pm0.041$        \\
$\Omega_{\rm P}$\tablenotemark{d}                & $6.09\pm0.11$          \\
$\Omega_{\rm S}$\tablenotemark{d}                & $4.64\pm0.03$          \\
\multicolumn{2}{c}{{\it Radial Velocity Curve Analysis}}                            \\
$K_{\rm P}$                                      & $205.2\pm5.1$ km~s$^{-1}$        \\
$K_{\rm S}$                                      & $174.2\pm5.2$ km~s$^{-1}$        \\
$q$\tablenotemark{e}                             & $1.178\pm0.046$                  \\
$v_{\gamma}$                                     & $261.4\pm4.6$ km~s$^{-1}$        \\
$a$                                              & $20.49\pm0.45$ R$_{\odot}$       \\
\multicolumn{2}{c}{\it Energy Distribution Analysis}                                \\
$T_{\rm eff}^{\rm S}$                            &  $20500\pm470$ K                 \\
${\rm[m/H]}_{\rm PS}$                            &  $-0.34\pm0.06$\phm{$-$}         \\
$\mu_{\rm PS}$                                   &  0 km s$^{-1}$                   \\
$E(B-V)$                                         &  $0.066\pm0.004$ mag             \\
$\log (R_{\rm S}/d)^2$                           &  $-23.007\pm0.023$ \\
\enddata
\tablenotetext{a}{Contribution of ``3rd light'' component to total observed flux, normalized at phase 0.25.} 
\tablenotetext{b}{Fractional stellar radius $r \equiv R/a$, where $R$ is the stellar ``volume radius'' and $a$ is the orbital semi-major axis.}
\tablenotetext{c}{Value and uncertainty computed directly from $r_{\rm P}$ and $r_{\rm S}$. A Monte Carlo simulation was run because of concerns with possible error correlations and yielded a very similar value of $<r_{\rm S}/r_{\rm P}>=1.566\pm0.036$.}
\tablenotetext{d}{Normalized potential at stellar surface.}
\tablenotetext{e}{Mass ratio ${M_{\rm S}}/{M_{\rm P}}$}.
\normalsize
\label{tabPARMS}
\end{deluxetable}
\normalsize


\begin{deluxetable}{ccc}
\tablewidth{0pc}
\tablecaption{Offsets Applied to HST/STIS Observations of EROS~1044}
\tablehead{
\colhead{HST Dataset}  &
\colhead{STIS Grating} &
\colhead{Offset}       \\
\colhead{Name}         &
\colhead{ }            &
\colhead{(${\rm FOS - STIS}$)} }
\startdata
O665A1030  & G430L & $+6.7\pm0.3$\%\\
O665A2030  & G750L & $+4.5\pm0.7$\%\\
\enddata
\label{tabSTIS}
\end{deluxetable}


\begin{deluxetable}{clc}
\tablewidth{0pc}
\tablecaption{Extinction Curve Parameters for EROS~1044}
\tablehead{
\colhead{Parameter}   &
\colhead{Description} &
\colhead{Value}        }
\startdata
$x_0$       & UV bump centroid        & $4.68\pm0.04$ ${\rm\mu m^{-1}}$\\
$\gamma$    & UV bump FWHM            & $0.79\pm0.14$ ${\rm\mu m^{-1}}$\\
$c_1$       & linear offset           & $\llap{$-$}0.35\pm0.29$        \\
$c_2$       & linear slope            & $0.89\pm0.08$          \\
$c_3$       & UV bump strength        & $0.69\pm0.26$          \\
$c_4$       & FUV curvature           & $0.46\pm 0.09$          \\
$R(V)$      & ${\rm A(V)/E(B-V)}$     & 3.1 (assumed)           \\
\enddata
\tablecomments{The extinction curve parameterization scheme is based on the work 
of Fitzpatrick \& Massa 1990 and the complete UV-through-IR curve is constructed 
following the recipe of Fitzpatrick 1999.}
\label{tabEXT}
\end{deluxetable}


\small
\begin{deluxetable}{lcc}
\tablewidth{0pc}
\tablecaption{Physical Properties of the EROS~1044 System}
\tablehead{
\colhead{Property}     &
\colhead{Primary}      &
\colhead{Secondary}    \\
\colhead{}             &
\colhead{Star}         &
\colhead{Star}          }
\startdata
Spectral Type\tablenotemark{a}       & B2 IV--V               & B2 III--IV               \\
$V$\tablenotemark{b} (mag)           & 16.68                  & 15.77                    \\
Mass\tablenotemark{c} (M$_{\sun}$)   & \phn$7.1\pm0.5$\phn    & \phn$8.4\pm0.5$\phn      \\
Radius\tablenotemark{d} (R$_{\sun}$) & $4.20\pm0.13$          & $6.61\pm0.14$            \\
$\log g$\tablenotemark{e} (cgs)      & $4.043\pm0.040$        & $3.721\pm0.030$          \\
T$_{\rm eff}$\tablenotemark{f} (K)   & $21400\pm525$\phn\phn  & $20500\pm470$\phn\phn    \\
$\log (L/L_{\sun})$\tablenotemark{g} & $3.53\pm0.05$          & $3.85\pm0.04$            \\
$[$m/H$]$\tablenotemark{h}          & $-0.34\pm0.06$\phm{$-$}& $-0.34\pm0.06$\phm{$-$}   \\
$v \sin i$\tablenotemark{i} (km s$^{-1}$) & $78$\phn          & $123$\phn                \\
$M_{bol}$\tablenotemark{j} (mag)     & $-4.07\pm0.13$\phm{$-$}& $-4.87\pm0.10$\phm{$-$}  \\
$M_{V}$\tablenotemark{k} (mag)       & $-1.90$                & $-2.82$                  \\
Age\tablenotemark{l} (Myr)           & \multicolumn{2}{c}{$29\pm4$}                      \\
$d_{\rm EROS~1044}$\tablenotemark{m} (kpc)& \multicolumn{2}{c}{$47.5\pm1.8$}             \\
\enddata
\tablenotetext{a}{Estimated from $T_{\rm eff}$ and $\log g$}
\tablenotetext{b}{From synthetic photometry of best-fitting {\it ATLAS9} model to the EROS~1044 
system. The ``3rd light'' component has a (synthetic) $V$ magnitude of 17.53.  Combining these magnitudes yields $V_{\rm EROS~1044} = 15.24$, consistent with the observational result from OGLE quoted in \S \ref{secOPHOT}.}
\tablenotetext{c}{From the mass ratio $q$ and the application of Kepler's Third Law.}
\tablenotetext{d}{Computed from the relative radii $r_{\rm P}$ and $r_{\rm S}$ and the orbital semimajor axis $a$.}
\tablenotetext{e}{Computed from $g = G M / R^2$}
\tablenotetext{f}{Direct result of the spectrophotometry analysis and photometrically-determined temperature ratio.}
\tablenotetext{g}{Computed from $L = 4 \pi R^2 \sigma T^4_{\rm eff}$}
\tablenotetext{h}{Direct result of the spectrophotometry analysis.}
\tablenotetext{i}{Computed assuming synchronous rotation.}
\tablenotetext{j}{Computed from $\log (L/L_{\sun})$ and $M_{bol\sun} = 4.75$}
\tablenotetext{k}{Computed from the synthetic V photometry, ${\rm E(B-V)}$, and distance.  Note that this result is consistent with expectations for mildly evolved early-B stars.}
\tablenotetext{l}{From the best-fitting isochrone to the data shown in Fig. \ref{figHRD}.}
\tablenotetext{m}{Using $(R_{\rm S}/d)^2$ from the spectrophotometry analysis and $R_{\rm S}$ from the light curve and radial velocity curve analyses. See \S 6.}
\label{tabSTAR}
\end{deluxetable}
\normalsize


\small
\begin{deluxetable}{llccc}
\tablewidth{0pc}
\tablecaption{Distances to LMC EB Systems}
\tablehead{
\colhead{}           &
\colhead{}           &
\colhead{}           &
\colhead{$\rm d_{LMC}$}      &
\colhead{$\rm d_{LMC}$}      \\
\colhead{EB System}   &
\colhead{Reference}   &
\colhead{$\rm d_{EB}$} &
\colhead{(Case I)\tablenotemark{a}} &
\colhead{(Case II)\tablenotemark{b}}}
\startdata
HV~2274    & Paper I,II & $47.0\pm2.2$ kpc & $45.9$ kpc & $47.0$ kpc \\
HV~982     & Paper II   & $50.2\pm1.2$ kpc & $50.6$ kpc & $50.7$ kpc \\
EROS~1044  & This Paper & $47.5\pm1.8$ kpc & $47.3$ kpc & $47.4$ kpc \\
\enddata
\tablenotetext{a}{Distance to a reference point at ($\alpha$,
$\delta$)$_{1950}$ = ($5^h 24^m$, $-69\arcdeg\/ 47\arcmin$),
corresponding to the optical center of the LMC's Bar according to
Isserstedt 1975.  Adopted LMC orientation defined by an inclination angle of
$38\arcdeg$ and a line-of-nodes position angle of $168\arcdeg$, from
Schmidt-Kaler \& Gochermann 1992.}
\tablenotetext{b}{Distance referred to the same reference point as
above.  Adopted LMC orientation defined by an inclination angle of
$34.7\arcdeg$ and a line-of-nodes position angle of $122.5\arcdeg$,
from van der Marel \& Cioni 2001.}
\label{tabDIST}
\end{deluxetable}


\end{document}